\documentclass[seceq,preprint]{ptptex}

\usepackage{xcolor}
\usepackage{graphicx}
\usepackage{wrapft}
\usepackage{epsfig}

\newcommand{\be}{\begin{equation}}
\newcommand{\ee}{\end{equation}}
\newcommand{\bea}{\begin{eqnarray}}
\newcommand{\eea}{\end{eqnarray}}

\usepackage[normalem]{ulem}

\begin{document}

\title{Modelling double charge exchange response function for tetraneutron system}


\author{R. \textsc{Lazauskas$^a$, J. \textsc{Carbonell}$^b$ and E. \textsc{Hiyama}$^c$}  }

\inst{
\\
$^a$IPHC, IN2P3-CNRS/Universite Louis Pasteur BP 28, F-67037 Strasbourg Cedex 2, France
\\
$^b$Institut de Physique Nucl\'eaire, Universit\'e Paris-Sud, IN2P3-CNRS, 91406 Orsay Cedex, France
$^c$RIKEN Nishina Center, 2-1 Hirosawa, Wako 351-0106, Japan
}



\abst{
This work is an attempt to model the $4n$ response function of a
recent RIKEN experimental study of the double charge exchange
$^4$He($^8$He,$^8$Be)$^4$n reaction in order to put in evidence an
eventual enhancement mechanism of the zero energy cross section,
including  a near-threshold resonance. This resonance can indeed
be reproduced only by adding to the standard nuclear Hamiltonian
an unphysically large T=3/2 attractive 3n-force which destroys the
neighboring nuclear chart. No other mechanisms like cusps or
related structures were found. }


\maketitle

\section{Introduction}

In a recent experiment at RIKEN \cite{Kisamori_PRL,Kisamori_PhD}
it was suggested  that the existence of a resonant tetraneutron
($4n$) state could explain the sharp structure observed in the
$^4$He($^8$He,$^8$Be)4n reaction cross section near the $4n$
threshold. They reported $E_R=0.86\pm0.65\pm1.25$ MeV and $\Gamma<
2.5$ MeV.

A subsequent theoretical analysis by the same authors
\cite{HLCK_PRC93_2016} of the present work, showed the difficulty
to accommodate such near-threshold resonance of the $4n$ system
without dramatically disturbing the well established neighboring
nuclear chart. It was however pointed out in
\cite{HLCK_PRC93_2016} that some reaction mechanism,  being able
to produce an enhancement of the cross section at small energy,
should be investigated. It is indeed well known that  without
presence of S-matrix poles there exist other possibilities to
generate sharp structures in a reaction cross
section~\cite{Calucci}.

The dineutron-dineutron correlation has sometimes been invoked as a possible enhancement mechanism,
due to the large value of the scattering length \cite{LF_PAN_2008,Bertulani_Nature}.
However previous calculations \cite{Lazauskas_PhD_2003,Lazauskas-4n,Bertulani_2003}  indicated that the
 interaction between two (artificially bound) di-neutron
was repulsive and so the probability to find four neutrons at the same point of the phase space is very weak.
A similar conclusion was reached   in the framework of the Effective Field Theories (EFT) for a more general case of
fermionic systems close to the unitary limit \cite{dd_EFT_Petrov_2004,dd_EFT_PLB_2017}.
Their conclusions are model independent and rely only in the fact that the fermion-fermion scattering length
is much larger than the interaction range, which is the case of the neutron-neutron system.
In view of these results, and contrary to some theoretical claims, it seems very unlikely that the tetraneutron
system could manifest a nearthreshold resonant state.

The aim of this short note is to investigate a particular reaction
mechanism, which can model the  double charge exchange reaction
and kinematics involved at the RIKEN experiment and could generate
any nearthreshold structure in the cross section.

This reaction mechanism is based on factorizing the transition amplitude into an spectator smooth part
 and a term involving the charge exchange
between the initial ($^4$He) and final ($4n$) states, which could be responsible for sharp structure
either due to a cusp or to a presence of a resonance.
The same mechanism, though with different dynamical contents in the initial and final states, was used
 in the numerical simulation underlying the analysis of  RIKEN result
\cite{Kisamori_PhD, Kisamori_PRL}  and can be of some help as a guide  for ongoing or future experiments.

We will detail in the next Section the reaction mechanism we have considered to model the RIKEN experiment.
The formalism allowing us to access to the response function, as well as the computational method we have used to obtain
the solution of the four-body Hamiltonian, will be sketched in Section 3.
Section 4 will be devoted to present our results and we will present our conclusions in a last Section.

\section{Reaction mechanism}\label{Reaction mechanism}

The experiment held by Kisamori et al. used 186 MeV/u $^8$He beam to bombard  $^4$He.
The reaction $^4{\rm He}(^8{\rm He},^8{\rm Be})^4n$ has been studied in a very particular
 kinematical conditions, where most of the kinetic energy of the  projectile has been transferred to $^8$Be nucleus.
 The decay products of $^8$Be, namely the two alpha particles, were detected in order to reconstruct the kinematics.
 Though an accurate description of this 12-nucleon reaction is far beyond the reach of our numerical tools, the particular
  kinematics employed in this
 experiment suggests to use approximate methods in order to estimate the possible  response.

 The principal reaction mechanism is  a double charge exchange with little energy transfer to the $^4{\rm He}$ nucleus target,
 which transforms it into a tetraneutron.
 The transition amplitude for such a process might thus be split  in two pieces
 \begin{equation}
A\approx\langle ^4n |\hat{O}_1|^4{\rm He}\rangle \langle ^8{\rm Be}
|\hat{O}_2|^8{\rm He}\rangle  \; \label{eq:amp},
\end{equation}
where $O_i$ are some transition operators.
These two factors correspond respectively to the ''fast''  process   $\langle ^8{\rm
Be}|\hat{O}_2|^8{\rm He}\rangle$ carrying  most of the 186 MeV/u kinetic energy of the projectile
and a ''slow'' one  $\langle^4n |\hat{O}_1|^4{\rm He}\rangle$  constituent of the charge exchange reactions and which remains practically static.

Total reaction cross section takes then the form:
\begin{equation}
\sigma_{tot}(E)\propto|\langle ^4n |\hat{O}_1|^4{\rm He}\rangle \langle ^8{\rm Be}
 |\hat{O}_2|^8{\rm He}\rangle |^2 \delta(E_i-E_f),
 \label{eq:cs}
\end{equation}
We are interested in the first term $\langle ^4n |\hat{O}_1|^4{\rm
He}\rangle$ of the last expression, since this term should bring
into  evidence any resonant features of the tetraneutron or any
alternative mechanism for enhancing the cross section (if at all).
The other term, related with a rapid process and involving large
momenta, may affect the overall size of the total cross section,
but should not have significant influence on the low-energy distribution of $^4n$ system.

On the other hand, the features of
$\langle ^4n |\hat{O}_1|^4{\rm He}\rangle$ matrix element will
critically depend on the particular transition operator
$\hat{O}_1$, which is unknown. In this work we will rely in
assuming  the most probable one. Since $^4{\rm He}$ and $^4n$ wave
functions are coupled with little momenta transfer, the
corresponding transition operator should contain only low order
momenta terms and thus its space-spin structure should have quite
a simple form. Furthermore, we will assume that both $^4$He and
$^4n$ wave functions are $J=0^+$ states since, as pointed out in
our previous studies \cite{Lazauskas-4n,HLCK_PRC93_2016}, this
state is the most favorable tetraneutron configuration  revealing
resonant features. The transition operator $\hat{O}_1$ should be
therefore  a scalar.

One possibility could be  $E_0$ or $\sigma_i.\sigma_j$ operators.
However the effect of these operators would be strongly suppressed
by the spatial orthogonality between the $^4{\rm He}$ and $^4n$
wave functions. This follows from the shell model representation
of $^4{\rm He}$ and $^4n$ wave functions with   s-wave protons
replaced by p-wave neutrons. Furthermore the 
{$\sigma_i.\sigma_j$ term}  implies correlated double-charge
exchange, but since exchange of the nucleons takes very short time
uncorrelated process is expected to dominate. The simplest
operator allowing such a transition might be represented as a
double spin-dipole term:
\begin{equation}
\hat{O}_1=(\sigma_i.r_i)(\sigma_j.r_j)\tau_i^-\tau_j^-
 \label{eq:op},
\end{equation}
In the last expression $\tau_i^-$ isospin reduction operators are
added which enable charge exchange, i.e. replace a proton by
neutron.

\bigskip
Once fixed the transition operator we are interested in evaluating the response (or strength) function, given  by
\begin{equation}\label{S_E}
S(E)=\sum_{\nu }\left\vert \left\langle \Psi _{\nu }\left\vert \widehat{O}_1\right\vert \Psi _{0}\right\rangle \right\vert ^{2}\delta (E-E_{\nu }),
\end{equation}
where $\Psi _{0}$ represents the ground state wave function of the $^4$He nucleus, with ground-state
energy $E_{0}$,  and  $\Psi_{\nu }$ represents the wave function of the $^4n$ system in the continuum with an
energy $E_{\nu }$.
Both wave functions are solutions of the four-nucleon Hamiltonian $H$. The energy is measured from some standard
value, e.g. a particle-decay threshold energy.

The Strength function (\ref{S_E}) may be rewritten  in terms of the forward propagator
\[\hat{G}^+(E ) = {1\over E- \hat{H}+i\epsilon} \]
in the following form
\begin{equation}
S(E) =-\frac{1}{\pi } \; {\rm Im} \; \left\langle \Psi _{0}\left\vert \widehat{O}_1^{\dag }\hat{G}^+(E )\widehat{O}_1\right\vert \Psi _{0}\right\rangle    \label{Strenght_func}
\end{equation}
in which the summation over the final states is avoided.

The later expression can still be transformed into

\begin{equation}
S(E)=-\frac{1}{\pi } \; {\rm Im} \left\langle \Psi _{0}\left\vert
\widehat{O}_1^{\dag }\right\vert \bar{\Phi} _{0 }(E)\right\rangle.
\label{Strenght_func}
\end{equation}
that is, as a matrix element of the of the (conjugate) transition
operator between the initial state $\Psi_0$ and the the  outgoing
collision wave-function $\bar{\Phi} _{0 }(E)$, defined as
\begin{equation}
\bar{\Phi} _{0 }(E)=\hat{G}^+(E )\widehat{O}_1\Psi _{0}
\end{equation}
Naturally, this wave function is a solution of the
inhomogeneous equation
\begin{equation}\label{Eq_Inho}
(E-\hat{H}+i\epsilon) \bar \Phi _{0}(E)=\widehat{O}_1\Psi _{0}.
\end{equation}
at a chosen energy E.

The right hand side of the former equation is compact, damped by the bound-state $%
\Psi _{0}$ wave function. The wave function $\bar\Phi _{0}$
asymptotically contains only outgoing waves. However asymptotic
structure of this wave function is rather complicated, involving
multidimensional four-neutron break up amplitudes. Nevertheless
the last inhomogeneous equation may be readily solved using
complex scaling techniques, we present in the next section.

\section{Computational techniques: Complex scaling Method}\label{Method}

To properly account for the boundary conditions of the resonance we have used the Complex Scaling
Method (CSM) \cite {Nuttal_PR188_1969,CSM-ref1,CSM-ref2}.
This computation technique, allowing to access the scattering solutions
with square integrable functions, was applied  to the response function
in\cite{Myo_PTP99_1998}.
Some recent applications and a more complete reference list can be found
in \cite{Aoyama_PTP116_2006, Horiouchi_PRC85_2012,Myo_2014}.

To this aim we have considered the four-body Hamiltonian $H$ in
configuration space and applied to each of the internal Jacobi
coordinates -- denoted generically by $X$ -- a complex rotation
with angle $\theta$, that is a mapping
\[ X\to X e^{i\theta} \]
The four-body Hamiltonian is transformed accordingly, as well as
the corresponding Schr\"{o}dinger equation
\[ H^{\theta} \Psi^{\theta} = E^{\theta}  \Psi^{\theta} \]
By doing so, and according to the  so-called ABC theorem \cite
{CSM-ref1,CSM-ref2}, the resonant poles are -- up to  numerical
inaccuracies -- independent of the parameter $\theta$ and are
isolated from the discretized non-resonant continuum spectrum,
provided some restrictions on the rotation angle are satisfied.

Let see see how the complex scaling method might be used to evaluate matrix element in  $(\ref{Strenght_func})$.
To this aim one must just apply the complex scaled expressions to both sides of the equation (\ref{Eq_Inho}), which become
\begin{equation}\label{Inho_CS}
(E-H^{\theta })\overline{\Phi}^{\theta } _{0}
(E)=\widehat{O}^{\theta }\Psi _{0}^{\theta }.
\end{equation}

The complex-scaled bound state wave function $\Psi_{0}^{\theta }$ is obtained by solving a bound state problem
with the complex-scaled Hamiltonian
\begin{equation}\label{He_CS}
(E_0-H^{\theta })\Psi _{0}^{\theta }=0,
\end{equation}
and it finally remains to compute the integral expression
\begin{eqnarray}
S(E) =-\frac{1}{\pi }Im\left\langle \widetilde{\Psi} _{0}^{\theta
}\left\vert (\widehat{O}_1^{\dag })^\theta\right\vert
\overline{\Phi} _{0 }^{\theta }(E)\right\rangle.
\end{eqnarray}
The solutions of the four-body equations (\ref{Inho_CS}) and
(\ref{He_CS})  have been obtained by using two different
approaches: the Faddeev-Yakubovsky equations in configuration
space and a variational Gaussian expansion method. The details of
this calculations and the numerical methods used can be found in
Ref. \citen{HLCK_PRC93_2016}.

\section{Results}\label{Results}

The nuclear Hamiltonian considered in our recent work
\cite{HLCK_PRC93_2016} consists in the Argonne AV8'  two-neutron
interaction \cite{AV18_1995}  plus three-nucleon forces in both
T=1/2 and T=3/2 total  isospin channels. The two-body and the
T=1/2 three-body parts were fixed in a previous work
\cite{Hiya04SECOND} in order to reproduce some selected  A=3 and
A=4 phenomenology and since, have been kept unchanged. The only
part of the Hamiltonian that was tuned in order to accommodate a
$4n$ resonant state, was the T=3/2 three-nucleon force. The latter
was chosen to have the same form than for the T=1/2 case, that is
a sum of two (attractive and repulsive) Gaussian terms:
\begin{equation}\label{V3NT}
V_{ijk}^{3N}=\!\!\sum_{T=1/2}^{3/2}\:
\sum_{n=1}^2 W_n(T)e^{-(r_{ij}^2+r_{jk}^2+r_{ki}^2)/b_n^2} \,
{\cal P}_{ijk}({T})\; .
\end{equation}
where ${\cal P}_{ijk}({T})$ is a projection operator on the total three-nucleon isospin $T$ state.

The parameters of this force are the following:
\begin{equation}\label{Para_T3/2}
\begin{array}{rclcl}
W_1(T=3/2)&=& \;\; {\rm free}, \;      &&  b_1=4.0\,\;  {\rm fm}, \cr
W_2(T=3/2)&=& +35.0 \;{\rm MeV},   &&  b_2=0.75\, {\rm fm} .
\end{array}
\end{equation}
They are all the same than for the T=1/2 except for the attractive
term $W_1(T=3/2)$  which is considered  as a free parameter, the
only one in our calculations. The model space of our calculations
was also identical to one used in a previous
study~\cite{HLCK_PRC93_2016}.  Namely for FY equations
partial-wave basis has been limited to angular momenta
max$(l,L,\lambda)\leq7$, providing total of 1541 partial
amplitudes. Furthermore $25^3$ Lagrange-mesh points were used to
describe radial dependence of Faddeev-Yakubovsky components,
resulting into linear-algebra problem of $2.4\times10^7$
equations. Such a large basis size ensured accurate results, which
can be traced by comparing FY calculation with Gaussian expansion
method in Table~\ref{Table_1}. Even for a very shallow
tetraneutron state of $\sim$1 MeV difference in calculated binding
energy was less than 20 keV, whereas expectation values differed
by less than 1\%. In the case of Gaussian Expansion Method, 
to obtain the converged energies, we included
angular momenta
max$(l,L,\lambda)\leq2$ and 14000 antisymmetrized four-body basis function.

 Our strategy to investigate the $4n$ resonant states
was quite simple: keeping unchanged the best established part of
the nuclear forces (two-body and T=1/2 three-body terms), we  have
first determined the strength of the three-nucleon force in the
T=3/2 channel which is required to reproduce the resonance
parameters given in Ref. \citen{Kisamori_PRL} and see then the
consequences of such a state in the nuclear chart.

\begin{figure}[h!]
\begin{minipage}[h]{7.5 cm}
\begin{center}
\epsfig{file=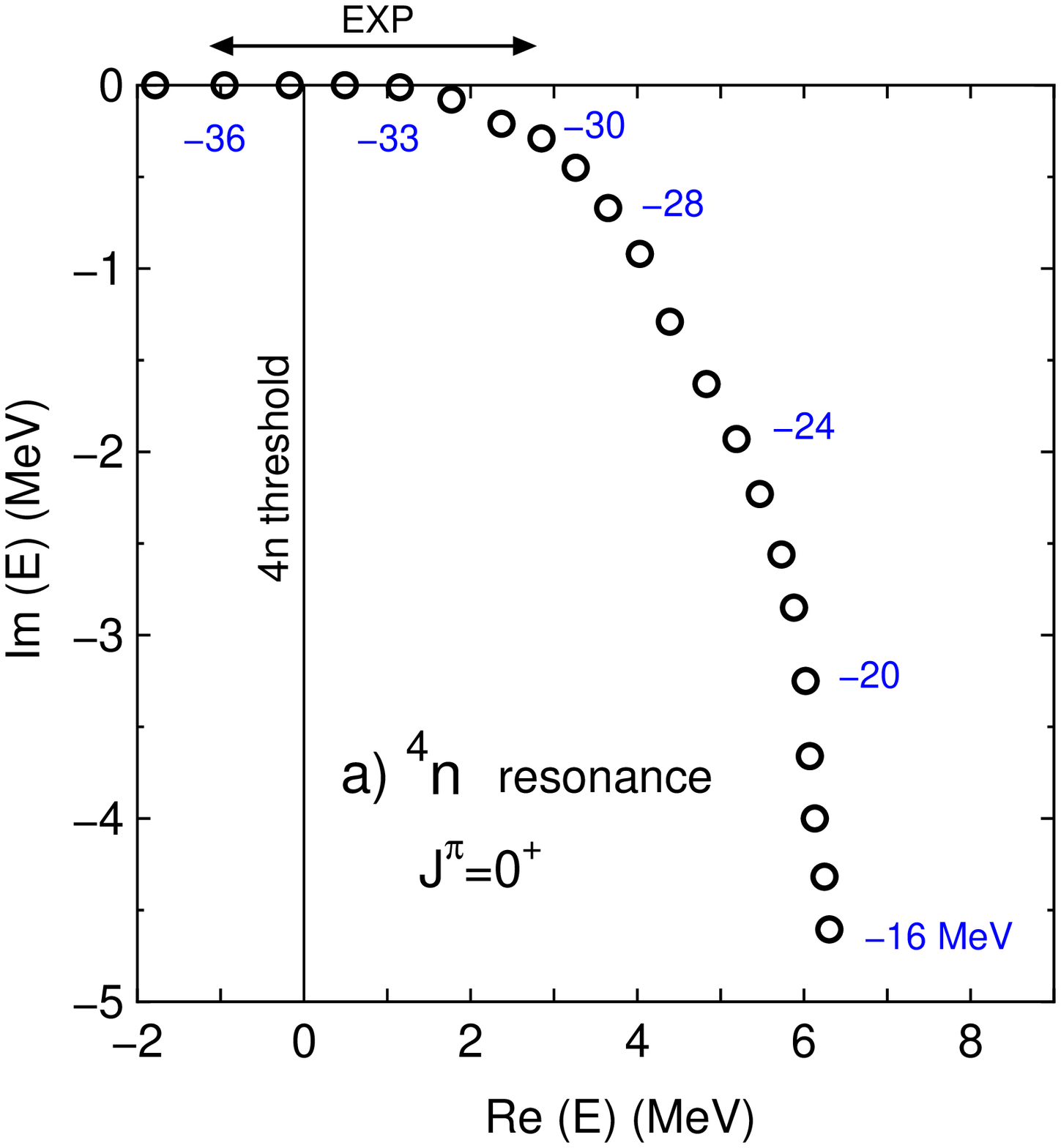,scale=0.35}
\caption{$4n$ resonance trajectory for $J^\pi=0^+$.  Strength
parameter $W_1(T=3/2)$   change from $-37$ to $-16$ MeV. The
parameters suggested in~\cite{Kisamori_PRL} is indicated by an
arrow.}\label{fig:nnnn-trajectry}
\end{center}
\end{minipage}
\hspace{.5cm}
\begin{minipage}[h]{7.5 cm}
\begin{center}
\epsfig{file=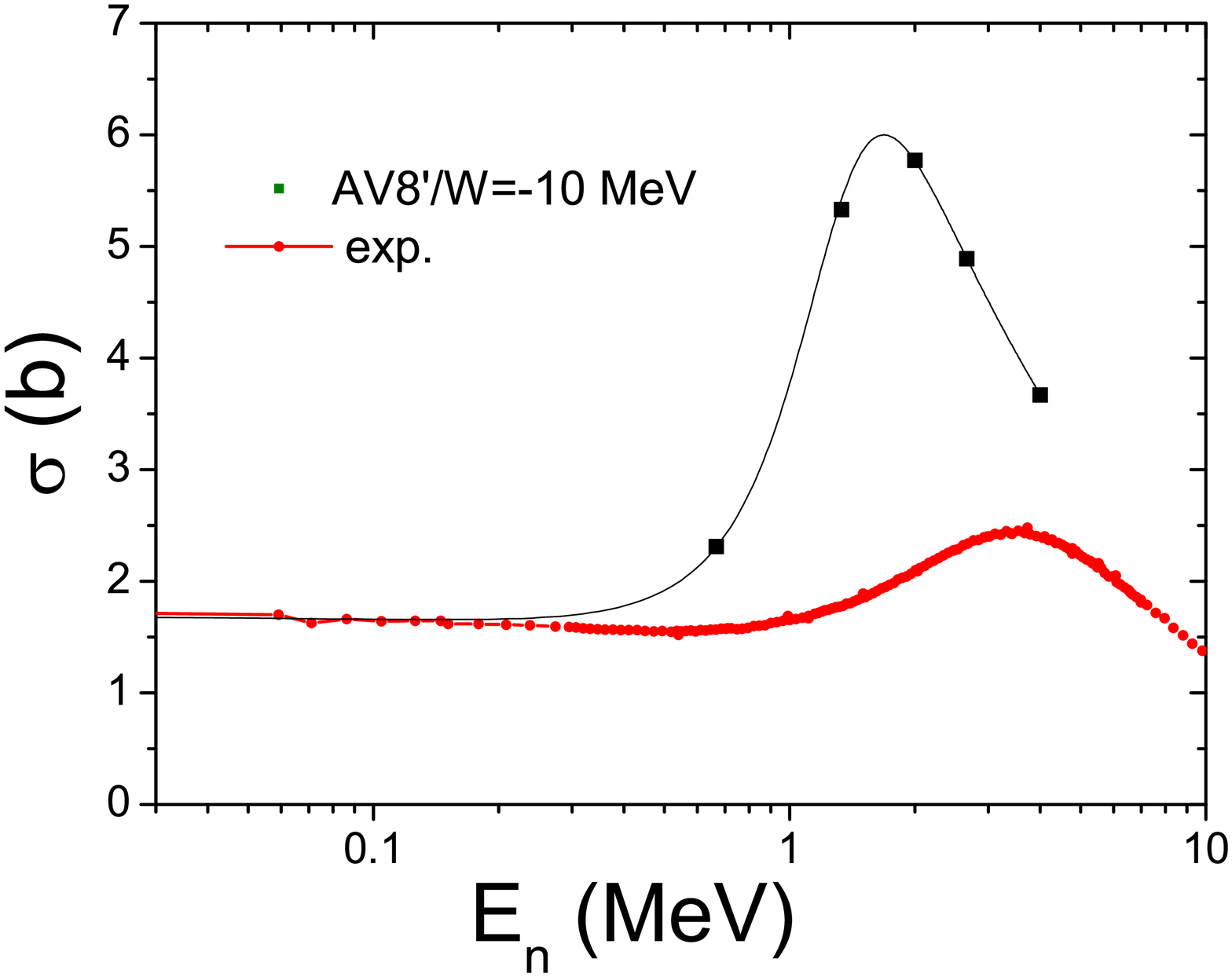,scale=0.33}
\caption{The computed $^3$H$+n$ elastic cross section (in black) using $W_1(T$=3/2)=$-10$ MeV is
compared to experimental data~\cite{Phillips-1980} (in red).}\label{fig:cross-section}
\end{center}
\end{minipage}
\end{figure}

In a first step, the critical $W_1(T=3/2)$ strength value at which
the $4n$ system is bound by $E=-1.07$ MeV, the lowest bound
compatible with the experimental values given in  Ref.
\citen{Kisamori_PRL}, was found to be $W^0_1(T=3/2)=-36$ MeV and
the most favorable state was $J{^\pi}=0^+$ followed by the
sequence  $J{^\pi}=2^+,1^+,2^-,1^-,0^-$ .

The strength parameter was then decreased in order to  move the bound
state state into the continuum and reproduce in this way  the observed resonance.
The $S-$matrix pole trajectory in the $4n$ complex energy plane was traced as a function of $W_1(T=3/2)$.
The results, taken from  Ref. \citen{HLCK_PRC93_2016}, are displayed in  Fig.~\ref{fig:nnnn-trajectry}.
As one can see, the range of $W_1$ values compatible with the experimental findings \cite{Kisamori_PRL},
indicated by an arrow in the upper part of the figure, are $W_1\in[-36,-30]$ MeV.

Several comments about this result are in order:

\begin{enumerate}
\item For comparison, the corresponding strength of the T=1/2
three nucleon forces is $W_1(T=1/2)=-2.04$ MeV and this allows to
increase the $^4$He binding energy by approximately 5 MeV. The
huge value, $W_1(T=3/2)\approx-30$ MeV,   required to generate a
$4n$ resonance, a factor $\sim15$ larger than for T=1/2, indicates
how deep in the continuum  the $4n$ resonant state should be
localized when we consider the standard nuclear Hamiltonian, that
is in particular with the two-neutron forces alone. Notice than
the pole trajectory displayed in Fig. \ref{fig:nnnn-trajectry} was
stopped at $W_1(T=3/2)=-16$ MeV. We have computed in  previous
work (see Figs. 3-6 in Ref.  \citen{Lazauskas-4n} ) the full
trajectory of the state until it is generated only by the
two-neutron forces and it turns to end in the third energy
quadrant. This feature is in sharp contrast with some recent works
of Refs. \citen{Pieper:2003dc,Andrei_2016,Gandolfi:2016bth}.
 The interactions  used by these authors are among the
best in the literature but we believe that the different
conclusions are due to the indirect approach they use to  deal
with the continuum in the four-body problem and to estimate the
$4n$  resonance positions.

\item The remarkably large value of the T=3/2 strength parameter  in the three-nucleon potential is  not understandable  in terms
of isospin symmetry of  nuclear force, which is quite accurately observed in phenomenology. It also  contradicts  the
 QCD inspired EFT  models
which found the T=3/2 contribution of three-nucleon force to be of subleading order
with respect to the T=1/2 ones \cite{VNNN}.
Any value of $\mid W_1(T=3/2)\mid $ larger  than  $\mid W_1(T=1/2)\mid$ is unphysical.

\item We have shown in Ref. \citen{HLCK_PRC93_2016} that for a
strength absolute value larger than 20 MeV the neighboring nuclei
$^4$H, $^4$Li and $^4$He (T=1) would be bound, contrary to the
well established experimental results. These nuclei  become
resonant only for $W_1(T=3/2)  \sim -20$ MeV, at which strength
the $4n$ system already correspond to a resonant state with
$E_R\sim 6$ MeV and $\Gamma\approx 7$ MeV. Still $W_1(T=3/2)  \sim
-20$ MeV  is an unphysical strength: even by taking half of it,
the n-$^3$H elastic cross section, displayed in Fig.
\ref{fig:cross-section}, would be in strong disagreement with the
experimental data.

\end{enumerate}

All the above discussed results led us to the conclusion
that the existence of a  $4n$  bound or low energy narrow resonant state is not compatible
with the well established facts of nuclear physics.
A similar conclusion is reached in a recent work \cite{Marek_4n_2016} using totally different
 interactions and techniques based on No-Core Gamow Shell Model
which takes properly into account the continuum.

Tetraneutron resonances certainly exists, even with pure
nucleon-nucleons forces, and we have computed them in a series of
works. They are however very far from the physical regions and
cannot manifest in a scattering experiment: any enhancement of the
reaction cross section involving $4n$ in the final state should
have an alternative dynamical explanation.

\begin{figure}[h!]
\begin{center}
\includegraphics[scale=0.5]{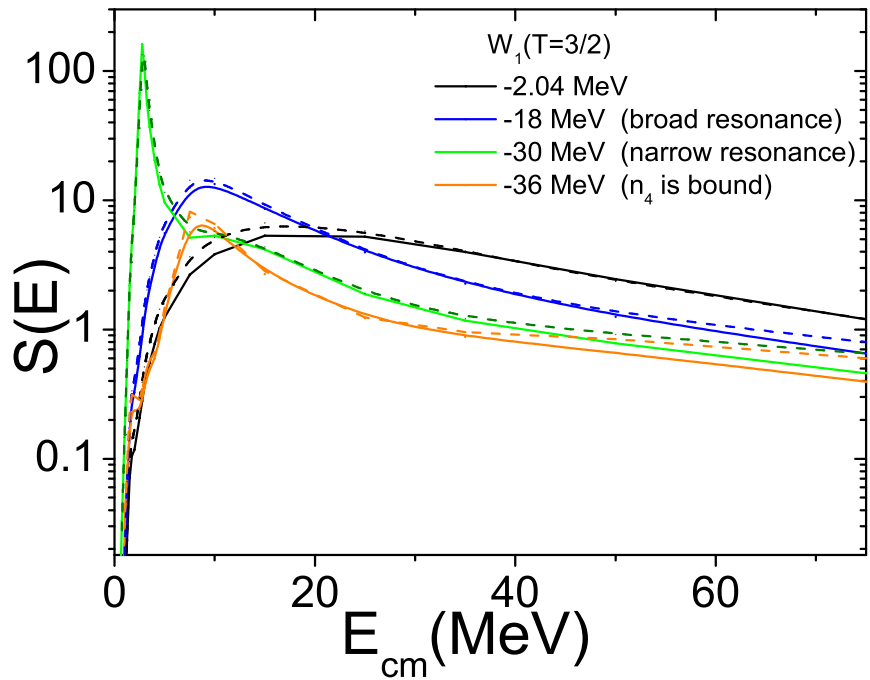}
\end{center} 
\vspace{.cm} \caption{ \label{fig:ptx} (Color online) Response
function for tetraneutron production from $\alpha$  particle due
to double-dipole charge exchange operator.}
\end{figure}

We will examine in what follows whether or not  the reaction mechanism described in the previous section
is able to produce any non resonant enhancement in the cross sections, as well as the consequences  that
 an eventual resonance could have on it.
The results are displayed in Fig.  \ref{fig:ptx}.

The black curve corresponds to the nuclear Hamiltonian, based on isospin independent three nucleon force.
In this case, the response function is flat without any near-threshold sharp structure.

By increasing the attractive part of the T=3/2 contribution, a
resonant peak appears. For $W_1(T=3/2) =-18$ MeV (blue curve),
still far from the values compatible with the RIKEN result,  the
underlying structure is already visible, although quite broad. It
becomes  sharper and sharper by further increasing the attraction
and moving the resonant pole close to the threshold.

For  $W_1(T=3/2)=-30$ MeV (green curve),  the tetraneutron resonance parameter  -- as given by
figure \ref{fig:nnnn-trajectry} -- are $E_R=2.8$ MeV  and $\Gamma=0.7$ MeV.
In the vicinity of this value
the corresponding response function  takes the usual Breit-Wigner form.

When further increasing the attraction the resonance becomes a
bound state (orange curve, corresponding to $W_1(T=3/2)=-36$ MeV
). The response function, which  has a pole at negative energy,
displays  also some pronounced structure at positive energies
although with reduced strength.

\begin{table} [htb] \begin{center}
\[\begin{array}{l c c c c c r } \hline \hline
\noalign{\vskip 0.15 true cm}
                    &    W_1(T=3/2)    &     E             &      <T>         &    <V_{2N}>     &      <V_{3N}>      &     {<V_{3N}>\over <V_{2N}>}   \\  
^4{\rm n}'     &     -36                 &   -1.00          &      67.02      &     -38.58         &         -29.52         &          76.5 \%  \\
                    &     -33                 &   +1.18        &     46.67       &     -28.13         &       -17.35           &          61.7  \%  \\
                    &     -30                 &   +2.70        &      29.11       &    -18.36         &          -8.05           &   43.8    \%  \\
                    &     -27                 &   +4.70        &     25.20       &  -15.03            &    -5.48                 &     36.5  \%  \\
                    &    -24                  &   +5.18        &    19.83        &  -11.98            &   -2.66                  &   22.2  \%  \\
\noalign{\vskip 0.15 true cm} \hline
                    &    W_1(T=3/2)    &     E              &      <T>         &    <V_{2N}>     &      <V_{3N}>      &     {\mid<V_{3N}>\mid\over \mid<V_{2N}>\mid}   \\  
                   & -36                &     -0.98          &  66.79          &           -38.47 &
                    -29.31          &  76.2\% \\
^4{\rm n}     &     -30                  & +2.84-0.33i  &     -               &   -26.7+6.5i    &  -10.1+4.4i          &   40.1 \%  \\
                   &    -24                   & +5.21-1.88i  &     -              &    -19.3+8.8i    &  -2.3+5.4i           &  27.7\%  \\
                     &    W_1(T=1/2)    &     E             &      <T>         &    <V_{2N}>     &      <V_{3N}>      &     {<V_{3N}>\over <V_{2N}>}   \\ \hline
^4{\rm He}    &    -2.04               &     -28.44     &     106.12     &    -131.17         &         -3.50          &                2.59 \%  \\
^4{\rm He*}   &                           &     - 8.13      &     49.36       &     - 56.71         &          -0.78         &       1.38 \%      \\
\noalign{\vskip 0.15 true cm} \hline
\end{array}\]
\end{center}
\caption{Two- and three-body contribution to the potential energy of the $4n$  system in a $J^{\pi}=0^+$ state as
a function  of $W_1(T=3/2)$ (all units are in MeV).
Results denoted by $^4$n' correspond to the bound state approximation and $^4$n to the continuum resonant states.
The results  are compared with the $^4$He ground and first excited state with the physical strength $W_1(T=1/2)=-2.04$.
The T=3/2 contribution in 4n required to accommodate a resonant state is more than one order of magnitude larger than
the T=1/2 (see rightest column).} \label{Table_1}
\end{table}

We would like to emphasize  that the results we have presented are
essentially independent of the nuclear Hamiltonian and the
mechanism considered to artificially produce the $4n$ bound or
resonant state. Several two- and three- and even four-nucleon
interactions have been indeed examined in  previous calculations
\cite{Lazauskas_PhD_2003, Lazauskas-4n,HLCK_PRC93_2016} and led to
very similar results. The underlying reason is  that, when any
ad-hoc mechanism is considered to enhance the 4n attraction in
order to accommodate a resonant state, this state is in fact,
essentially supported by the artificial  binding mechanism
adjusted to this aim: the details of the remaining nucleon-nucleon
interaction are residual.

This fact is  illustrated in Table \ref{Table_1} where we have
compared the contributions of the two- and three-nucleon force
(averaged values of the corresponding potential energies) both for
the $^4$He and the 4n system, for several values of the strength
parameter $W_1(T)$. As one can see from the results of this Table
the $V_{2n}$ and $V_{3n}$, the contributions to the $^4$n state in
the resonance region  are of the same order and its ratio (the
rightest column) remains in any case more than one order of
magnitude larger than for the T=1/2 case in $^4$He, the contrary
of one could expect from physical arguments.

 As it was pointed out in the Introduction the
dineutron-dineutron interaction is repulsive. This repulsion
relies on  very general arguments and any attempt  to bring
together four neutrons on a nearthreshold narrow state can only
come from an artificially ad-hoc extrabinding.

\section{Conclusion}

Inside the  simplistic reaction mechanism we have considered in
this paper, we  are not able to generate  an increasing of the
cross section at the origin other  than by  accommodating a sharp
$4n$ resonance. No other  possibilities like cusp or related
structures could have been  exhibited.

We have found in our previous work \cite{HLCK_PRC93_2016} that the existence of such
a resonance is hardly compatible with the well established properties of nuclear interactions
and experimental data on neutron rich nuclei. This is in agreement with the findings of
Ref. \citen{Marek_4n_2016}.
Its is worth noticing, however, that opposite conclusions have been reached in recent calculations
\cite{Andrei_2016,Gandolfi:2016bth}.

We believe that the reason for such a striking difference is not  the neutron-neutron interaction but rather the
approximate methods they use to deal with the 4n continuum.

The RIKEN $^4$He($^8$He,$^8$Be)$4n$ experiment \cite{Kisamori_PRL} has been repeated  with improved
statistics and is under analysis.
Other experiments are scheduled  at the same laboratory on $^8{\rm He}
 (p,p\alpha)4n$ \cite{RIBF}  and at  J-PARC
\cite{JPARC_2017}  on $^4{\rm He}(\pi^-,\pi^+)^4n$. We hope they will be decisive to clarify such
a challenging problem.

\section*{Acknowledgements}
The authors thank Dr. S. Shimoura,  Dr. M. Marques and Dr. N. Orr
for valuable discussion. The numerical calculation were performed
on the IDRIS Computer Center. Part of this results were obtained
during the Espace de Structure Nucl\'eaire Th\'eorique  (ESNT,
http://esnt.cea.fr) workshops at CEA from which the authors could
benefit. This work was partly supported by RIKEN iTHES Project and
France-Japan PICS.
In addition, this work was supported by JSPS Japan-France Joint Research Project
and IN2P3 project "Neutron-rich unstable light nuclei".



\end{document}